\documentclass[apjl]{emulateapj}

\usepackage[english]{babel}
\usepackage[utf8x]{inputenc}
\usepackage[T1]{fontenc}

\usepackage{amsmath}
\usepackage{amssymb}
\usepackage{graphicx}
\usepackage[colorinlistoftodos]{todonotes}
\usepackage[colorlinks=true, allcolors=blue]{hyperref}

\def\deg{\ifmmode^\circ\else$^\circ$\fi}
\def\arcsec{\ifmmode^{\prime\prime}\else$^{\prime\prime}$\fi}
\def\arcmin{\ifmmode^{\prime}\else$^{\prime}$\fi}

\newcommand{\OBJECT}{`Oumuamua }

\shorttitle{Interstellar Interlopers}
\shortauthors{Do, Tucker and Tonry}

\begin{document}

\title{Interstellar Interlopers: Number Density and Origin of `OUMUAMUA-like Objects}

\author{Aaron Do}

\author{Michael A. Tucker}

\author{John Tonry}
\affil{Institute for Astronomy \\
University of Hawai`i \\
2680 Woodlawn Dr. \\
Honolulu, HI 96822, USA}

\begin{abstract}
We provide a calculation of Pan-STARRS' ability to detect objects similar to the interstellar object 1I/2017 U1 (hereafter `Oumuamua), including the most detectable approach vectors and the effect of object size on detection efficiency. Using our updated detection cross-section, we infer an interstellar number density of such objects ($n_{IS} \approx 0.2~\rm{au}^{-3}$). This translates to a mass density of $\rho_{IS} \approx 4M_\oplus~\rm{pc}^{-3}$ which cannot be populated unless every star is contributing. We find that given current models, such a number density cannot arise from the ejection of inner solar system material during planet formation. We note that a stellar system's Oort cloud will be released after a star's main sequence life time and may provide enough material to obtain the observed density. The challenge is that Oort cloud bodies are icy and \OBJECT was observed to be dry which necessitates a crust generation mechanism. 
\end{abstract}

\section{Introduction}

\OBJECT was discovered by the Panoramic Survey Telescope and Rapid Response System \citep[Pan-STARRS, ][]{chambers2016} and was found to have a hyperbolic orbit \citep{2017MPEC....V...17W} with an eccentricity $e = 1.1956 \pm 0.0006$ \citep{Meech}. Its incident velocity is too high to have originated as a cometary object deflected by close passage to a giant planet \citep{2017RNAAS...1....5D} indicating \OBJECT is truly an interstellar object. The object's albedo and physical size give an absolute magnitude of $H \sim 22.1$. Light curve analysis suggests an effective spherical radius of $102\pm 4\,\rm{m}$, an axial ratio of $\gtrsim 10:1$, and little to no cometary activity \citep{Meech}.

\section{Inferred Number Density of such objects}
\label{sec:density}

We
estimated the total volume from which Pan-STARRS could have seen \OBJECT\
during a typical operations period of MJD
58000--58079 (\OBJECT\ was discovered on 58045).  During this 80 day period
6600 ``quad'' (4$\times$ revisit) observations were examined for moving objects.

The calculation adopts
an $H$ magnitude of 22.1 \footnote{\citet{Meech} record a median $H$ magnitude of 22.4 over a full rotational period. We use the detection magnitude from \url{https://ssd.jpl.nasa.gov/horizons.cgi}.}, a velocity at infinity of 26~km/s, and
a nominal phase function with slope parameter $G=0.15$. The
asteroid is allowed to have come from any direction and from both sides of the Sun,
sampling every 9~deg$^2$ of the sky.  For each quad observation
and direction of approach, we compute the minimum and maximum
distance that Pan-STARRS could have seen such an asteroid, and assume 
each Pan-STARRS observation covers 6~deg$^2$. The tangential
velocity with respect to the Earth causes substantial trailing losses.
We optimistically assume that the detections are complete to the limiting
magnitude, degraded by the additional background noise and the streak length,
and cut off for angular velocities greater than 10~deg/day.

\begin{figure}
\begin{center}
\includegraphics[width=\linewidth]{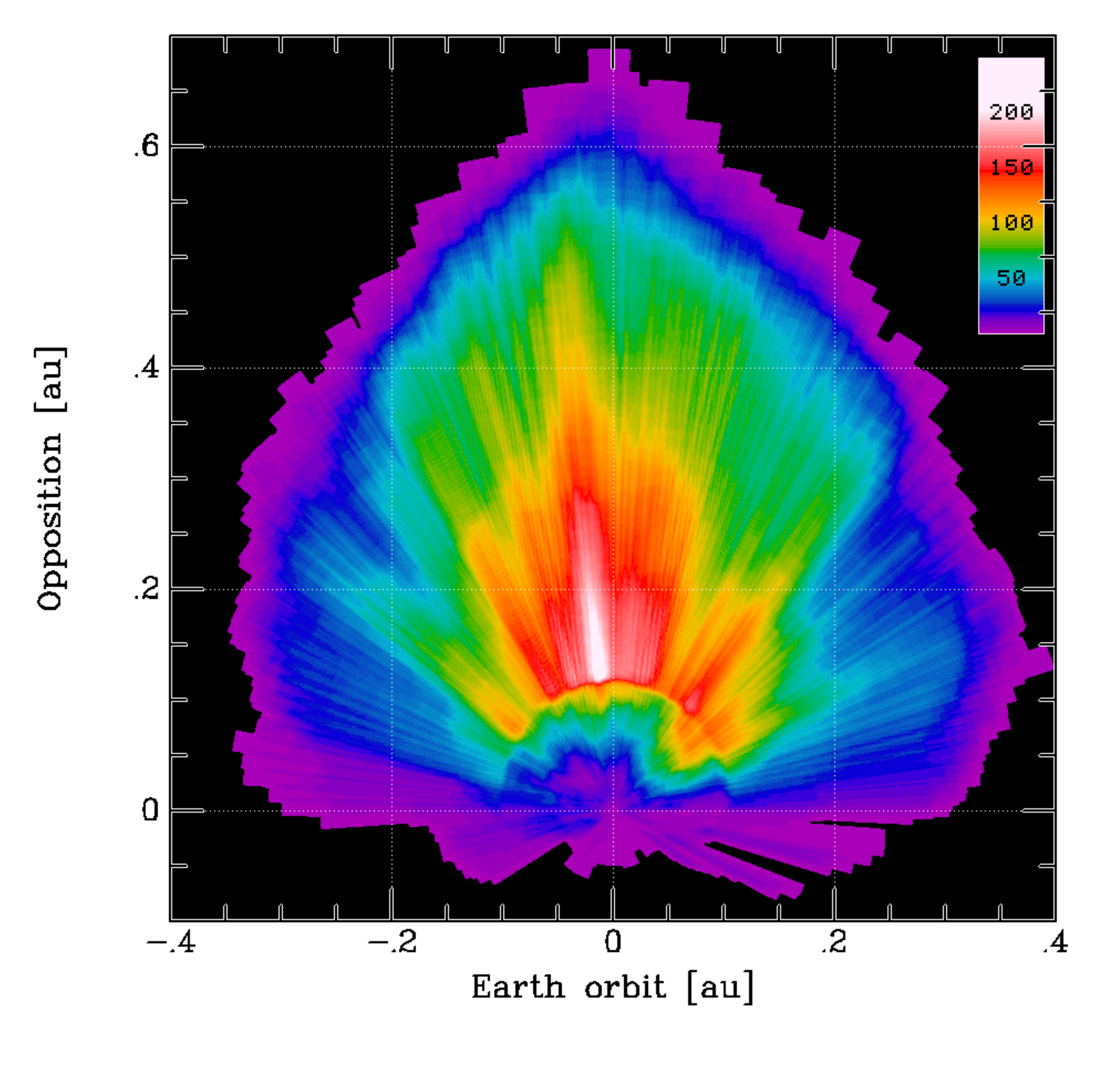}
\end{center}
\caption{The 6600 Pan-STARRS quad exposures obtained between MJD 58000 and
  58079 are projected in ecliptic $z$ onto Earth-orbit and Sun-opposition
  coordinates. The Sun is at (0,-1); color codes the number of visits.  
  The domain of sensitivity forms a ``candle flame'' away from the Sun because
  of the phase function for asteroid illumination and reflection.  The minimum and maximum
  distances are set by the tangential velocity and limiting magnitude,
  averaged over all possible directions of approach.}
\label{fig:flame}
\end{figure}

We then project this truncated
cone back to infinity along the assumed orbit and average the
resulting volume over all approach directions.  Figure~\ref{fig:flame}
illustrates where Pan-STARRS could have detected a similar interstellar object.

The Pan-STARRS detection process is not 100\% efficient over
the full 6~deg$^2$ and streak length of $\sim$20~arcsec, resulting in a
smaller surveyed volume and a larger inferred number density.
Losses include failure to trigger on a streaked object,
streaks that encounter "chip gaps" and do not generate enough detections
to be flagged, failure to link detections, and insufficient detections
and follow-up to determine an orbit. The limit of 10~deg/day used in
the calculation above is the limit for candidate detections to be linked,
for example, and the use of 6~deg$^2$ instead of the 7~deg$^2$ of
detector area attempts to address the loss from "chip gaps".

In order to explore the detection efficiency, we re-ran the simulation
with a limit of 5~deg/day and found a volume that is 0.75 of the 10~deg/day limit.
When we use a 10~deg/day limit but require that at least one spot on a streak
statistically deviate above the limiting point source magnitude, the calculated
volume is 0.65 of nominal. A proper assessment would require Pan-STARRS to do a
detailed Monte Carlo study of simulated streaks inserted into raw images and
probability of detection. For the purposes of this study we simply caution
that we are probably over-estimating the detection efficiency by a factor
of 1.3-2 and therefore under-estimating the density of interstellar objects.
However, we regard the statistical uncertainties from a single detection in
3.5 years to be more significant.

The outer contour of the ``candle flame'' in which Pan-STARRS is sensitive
to an object of this size has a cross section that is $\sim$4 times
larger than obtained by this projected volume calculation.
Pan-STARRS loses efficiency 
because not all exposures have full sensitivity and also because Pan-STARRS
revisits the entire volume about once per month, but the crossing time
for a typical trajectory is about 10 days.

A number of interesting results emerge from this calculation.

\begin{figure}
\begin{center}
\includegraphics[width=\linewidth]{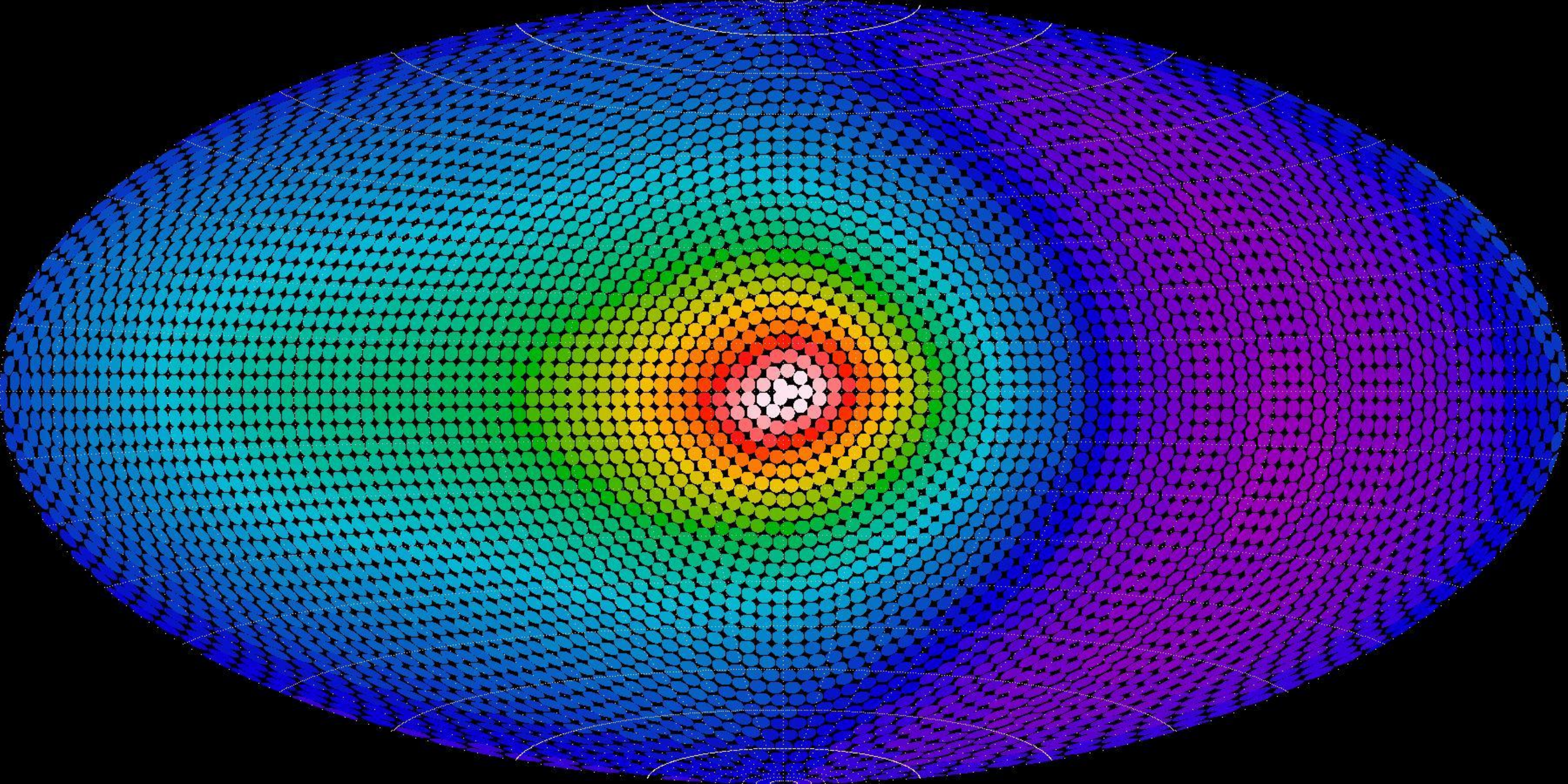}
\end{center}
\caption{The interstellar volume to which Pan-STARRS is sensitive is shown as a function of
  approach direction from infinity.  These are solar opposition coordinates, i.e. relative to
  the direction away from the Sun from Earth at the moment the asteroid passes into visibility.
  (Note that this shows original direction of approach, not the direction of travel when it is
  visible.) Horizontal is 0--360$^\circ$ from opposition in the plane of the ecliptic, vertical
  is $\pm90^\circ$ in ecliptic latitude.  The colors encode the logarithm of the survey volume,
  factors of 3 from magenta to green to red.}
\label{fig:skyvol}
\end{figure}

The most detectable direction
of approach of an asteroid is directly from
behind the Sun; ground-based surveys are
unlikely to see an interstellar object on approach.  Figure~\ref{fig:skyvol} shows the distribution of
detection volume as a function of initial approach angle.
The most probable direction of approach
is from the direction of the Sun because that direction has a bigger cross section
for asteroids to bend around and pass into the visibility volume (gravitational focusing).
Values are enhanced to the east because approach from that direction tends to align with the Earth's 
orbital velocity when the orbit passes into the Pan-STARRS detection volume.
    
The detection volume scales as $v_\infty^{-1}$ ($v_\infty^{-2}$
from gravitational focusing and $v_\infty^{+1}$ from 
impingement rate).  If there were a significant population of interstellar
objects with a velocity near that of the Sun they would be much more detectable than \OBJECT, so the velocity distribution of interstellar objects is more likely to
reflect the local standard of rest rather than the Sun's motion.

The detection volume scales crudely as the cube
of the size of the object, but with
significant deviations.  It increases as $D^2$ at kilometer sizes
because the Pan-STARRS
visibility stretches to distances where the Sun's illumination diminishes,
and it goes as $D^{4.5}$ at small sizes because the angular velocity
is too high at the close distances where the asteroid is detectable.
The Pan-STARRS volume surveyed per year is given by
\begin{equation}
V(x) = 4~\hbox{au$^3$ yr$^{-1}$}\;x^{4.5}\;(1+x^{1.67})^{-1.5},
\label{eq:vofd}
\end{equation}
where
\begin{equation}
x=10^{0.2(22.1-H)}.
\end{equation}

\begin{figure}
\begin{center}
\includegraphics[width=\linewidth]{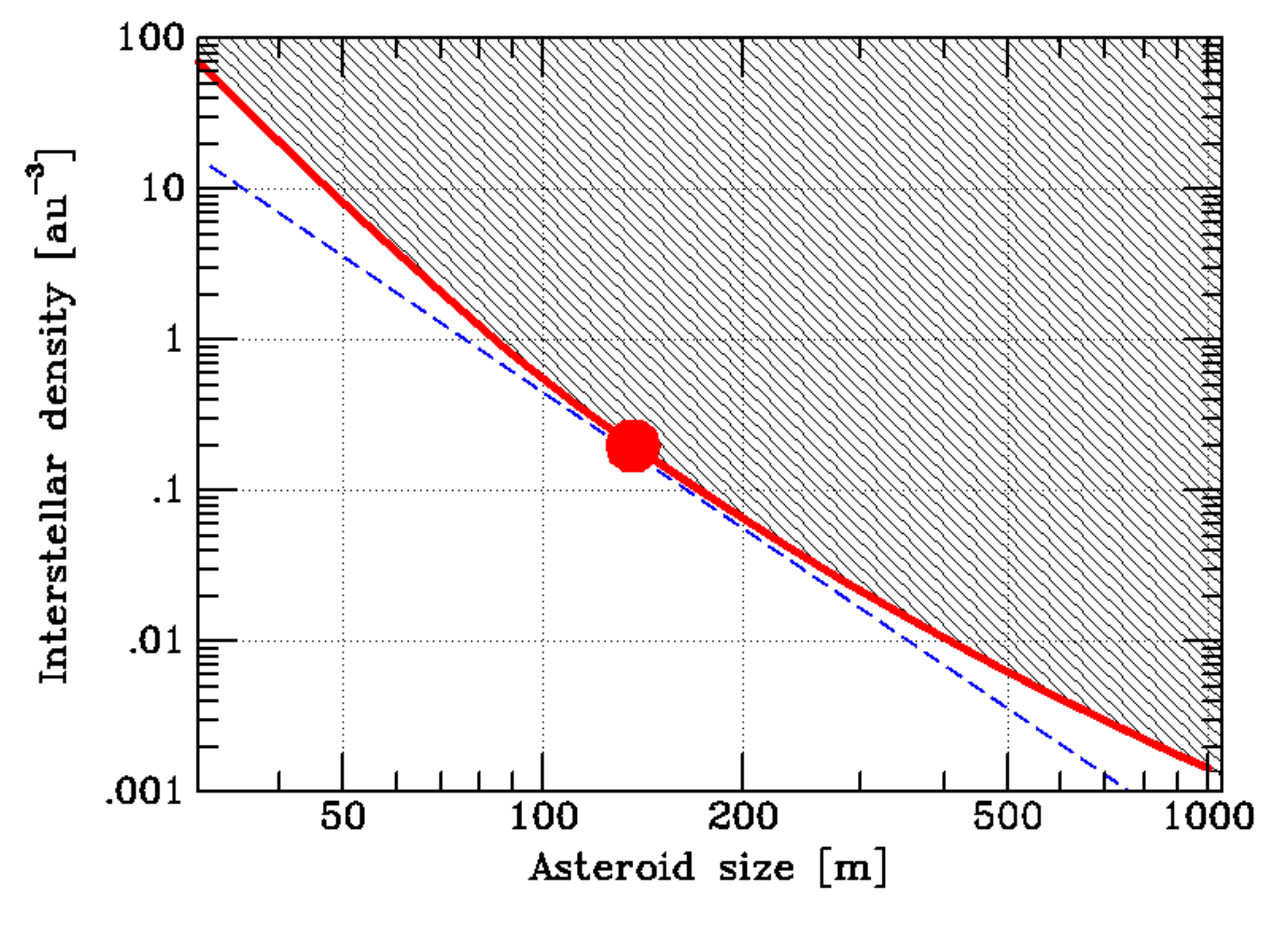}
\end{center}
\caption{Limits on the cumulative interstellar number volume of interstellar objects
  from the Pan-STARRS survey is shown as a function of object diameter (using a nominal
  $H=17.75$ and albedo 0.14 for 1~km to convert observable $H$ magnitude to diameter).  The red curve is the
  inverse of the total volume surveyed by Pan-STARRS and the dot shows the detection.  The actual distribution must also lie below the dashed blue curve which shows the
  $n(d>D)\propto D^{-3}$ where total mass diverges logarithmically.}
\label{fig:size}
\end{figure}

Detection rate divided by the rate that a survey covers volume is a measure of the cumulative number density of interstellar objects down to the detection diameter $D_0$, $n(d>D_0)$.
Figure~\ref{fig:size} shows the inverse of the total volume surveyed by Pan-STARRS during its 3.5 year mission to illustrate the estimated cumulative number density of
interstellar objects at least as large as \OBJECT.

Using inverse volume for the cumulative number density is an underestimate that depends on how steeply the cumulative distribution falls at large sizes. The cumulative distribution for the number density of interstellar objects as a function of size must be steeper than $n(d>D)\propto D^{-3}$ since the total mass of cannot diverge at large sizes. Inverse volume underestimates the true cumulative number by a factor of $(4-\alpha)/(2-\alpha)$ for $n(d>D)\propto D^\alpha$, so the inverse volume is at most a 40\% underestimate.  Because of the
curvature of the survey volume as a function of size, for any broad, smooth distribution \OBJECT\ has the most probable size that Pan-STARRS could detect.

The total volume for 6600 quads and 80 days is 0.3~au$^{3}$.  Since
Pan-STARRS has been operating in this search mode for 3.5 years, this
implies a total survey volume of 5~au$^3$ at this size. One detection over this time implies a mean density of 0.2~au$^{-3}$
or $\sim 2\times10^{15}$~pc$^{-3}$.

\section{Implications of Number Density}
\label{sec:implications}

The interstellar objects may come from a combination of different sources. The total number of ejected bodies required of a single source $N_{ej}$ and the corresponding ejected mass $M_{ej}$ are smallest when $n_{src}$ is as large as possible. As an upper limit we consider all stars as contributors to the density of interstellar objects. Assuming a stellar number density of $\sim 0.1~\rm{pc}^{-3}$ each star must account for over $10^{16}$ ejected bodies. 

We can also formulate the amount of ejected material as a total amount of mass ejected with some assumptions of the object's size and density. The total amount of ejected mass per system will be inversely proportional to the number density of the progenitors $n_{\rm{src}}$,
and directly proportional to the density of material $\rho$.
With a precise size distribution we would be able to calculate the aggregate volume of interstellar objects based on the abundance of $\sim 100~\rm m$ objects. However, with only a single detection the size distribution is undetermined. For an order of magnitude estimate, we assume the volume of \OBJECT is average for the interstellar population (effective spherical radius of 102 m). Combining these parameters,
we arrive at an equation for the amount of mass each system must eject to produce the calculated number density of \OBJECT like objects $n_{IS}$

\begin{multline}
\label{eq:m_ej}
M_{ej} = 40 M_\oplus (\frac{n_{IS}}{0.2~\rm{au}^{-3}}) (\frac{n_{src}}{0.1\,\rm{pc}^{-3}})^{-1} \\ (\frac{\rho}{3\, \rm g\,\rm{cm}^{-3}})(\frac{R}{102~\rm{m}})^3.
\end{multline}

\noindent We are implicitly assuming an isotropic and homogeneous distribution of objects.
In \S\ref{sec:planetformation} \& \ref{sec:postMS} we review possible ejection mechanisms and their effect on Eq. \ref{eq:m_ej}.

\subsection{Scenario I: Mass Ejection During Planet Formation}
\label{sec:planetformation}

Rocky objects similar to \OBJECT are thought to be ejected from the inner solar system in the early stages of the planet formation process \citep{2017RNAAS...1...13G, 2017arXiv171103558P}. Since protoplanetary disks can comprise a few percent of the star's mass \citep{2011ARA&A..49...67W}, our figure of $\sim 5~M_{\oplus}$ seems believable. However, several considerations undermine the feasibility of populating the interstellar number density of \OBJECT like objects through this method:
\begin{itemize}
  \item{} The significant driver for disk mass loss
   is photoevaporation \citep{2014prpl.conf..475A} whereas objects similar to \OBJECT 
   have to be ejected through gravitational processes (close encounters or orbital 
   resonances).
   The figure of merit when considering mass ejection during planet 
   formation is not protoplanetary disk mass but planetesimal disk mass. Estimates of the mass of 
   our solar system's planetesimal disk vary depending on the model: $\sim 14-28~M_\oplus$ from \citet{2013ApJ...768...45N}, $\sim 12-34~M_\oplus$ from \citet{2014ApJ...792..127R}, $\sim 13-65~M_\oplus$ from \citet{2017AJ....153..153D}.
  \item{} Gravitational processes are only efficient at mass ejection in
   systems with giant planets \citep{2011ARA&A..49..195A}.
   Observations place the abundance of star systems hosting at least one Jupiter-sized or larger 
   planet at $f_d \sim 0.1$ \citep{2016ApJ...819...28W} where $f_d$ is the fraction of systems that are considered ``disruptable''. Multi-star systems and close stellar passages 
   in clusters may increase the abundance of disruptable systems but likely not to more than
   $f_d = 0.15$, reducing $n_{src}$ proportionally.
   \item{} The entirety of the planetesimal disk will not be ejected into the ISM during this process. Some of the mass will remain behind to form asteroid belts, terrestrial planets, or outer solar system structures. With this in mind we take the ejected fraction of planetesimal material to have a fiducial value of $f_e \sim 0.8$.
\end{itemize}

Accounting for the reduced abundance of disruptable systems and the imperfect ejection efficiency, the necessary mass of the planetesimal disk approaches
\begin{equation}
M_{pd} \sim 320~M_\oplus (\frac{M_{ej}}{40 M_\oplus})(\frac{f_d}{0.15})^{-1} (\frac{f_e}{0.8})^{-1}.
\end{equation}
which exceeds all estimates of our solar system's planetesimal disk by an order of magnitude. Unless the planetesimal disk mass grows faster than the initial mass function, more massive stars having more massive planetesimal disks do not alleviate this issue. Mass ejection during planet formation seems insufficient to populate the density of \OBJECT like objects. 

\subsection{Scenario II: Mass Ejection By Oort Cloud Freeing}
\label{sec:postMS}

In addition to the planet formation process, stars also eject mass into the ISM during the transition off of the main-sequence \citep{2017arXiv171109599R}.

Larger stars ending in core-collapse supernovae over half their mass in a relatively short time span, liberating or vaporizing all bound objects. Kuiper belt objects are likely destroyed in the supernova but Oort cloud objects are distant enough to survive. Exposure to the supernova will remove volatiles from the surface of these Oort cloud objects and release them into the surrounding ISM. However, supernovae are rare, so even if an entire Oort cloud is dried and released with every supernova this method falls short of matching the observed interstellar number density.

\begin{figure}
\includegraphics[width=\linewidth]{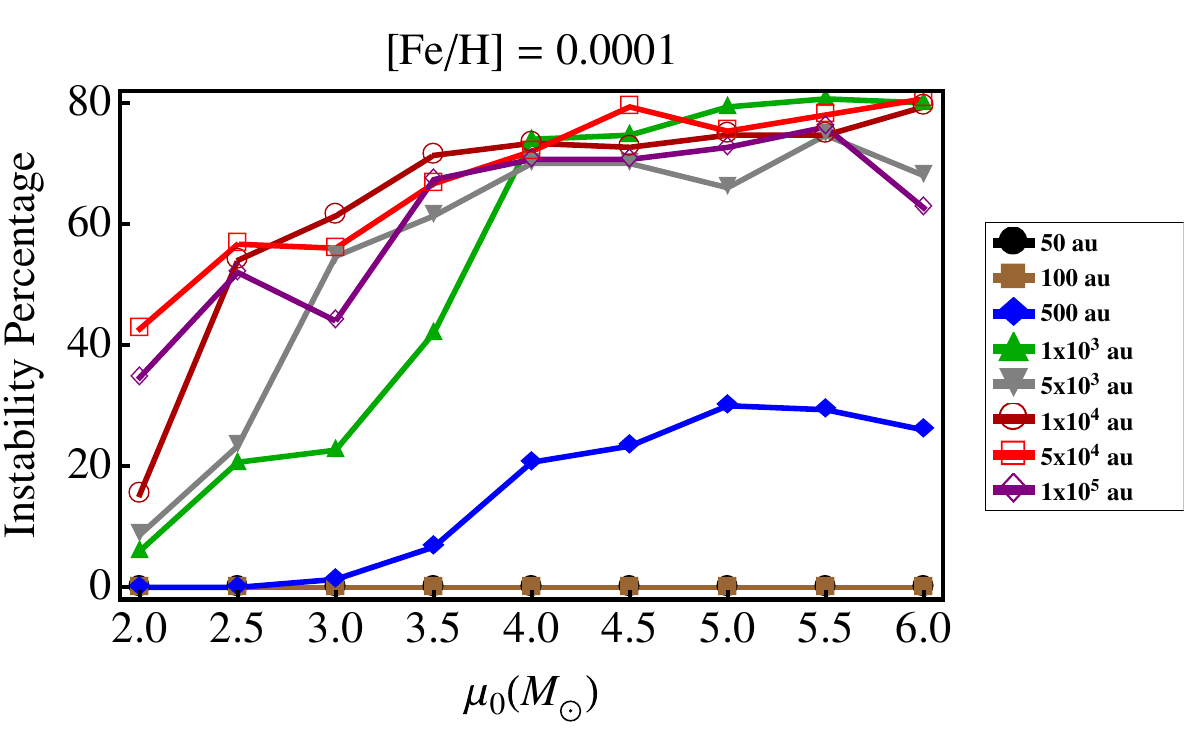}
\includegraphics[width=\linewidth]{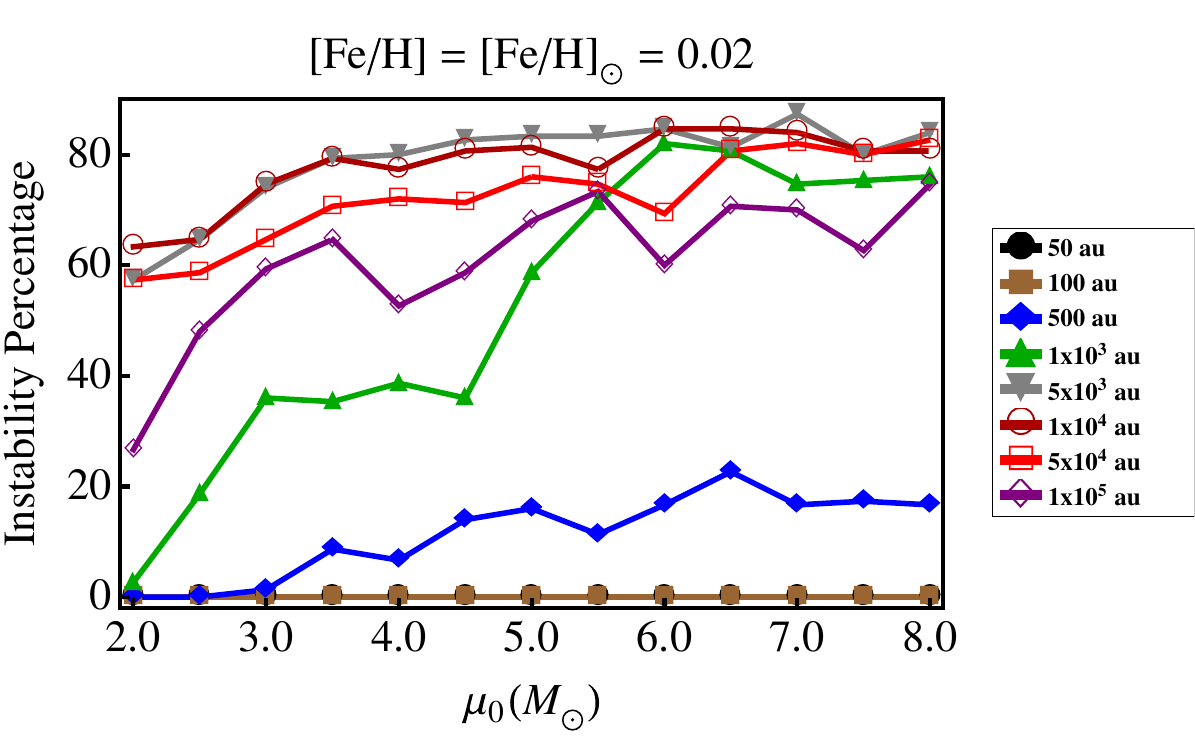}
\caption{From \protect\cite{2011MNRAS.417.2104V}, planetary ejection prospects for massive stars from $2M_{\odot} - 8M_{\odot}$ derived from multiple numerical simulations of low and solar metallicity host stars.}
\label{fig:instability}
\end{figure}

Considering intermediate mass stars with $2 - 8 M_\odot$, the transition though the asymptotic giant branch (AGB) phase also results in significant mass-loss. The mass-loss rate is insufficient for the release of all circular orbits, however, a large fraction of Oort cloud objects are released \citep{2011MNRAS.417.2104V}. Fig. \ref{fig:instability} from \citet{2011MNRAS.417.2104V} illustrates the instability percentage for objects at various orbital distances around intermediate-mass stars.

The Oort cloud's number density and size distribution have been topics of debate for as long as the Oort cloud hypothesis has existed \citep[for reviews see ][]{2004ASPC..323..371D, MAPS:MAPS12080, 2015SSRv..197..191D}. Direct observation of objects in the proposed Oort cloud region is difficult due to the distances involved. Comets injected from the Oort cloud into the inner solar system by gravitational perturbations imply a reservoir of some $\sim 10^{12}$ $\sim 2~\rm{km}$ sized objects in the inner Oort cloud \citep[$3~000~\rm{au} < a < 20~000~\rm{au}$, ][]{2009Sci...325.1234K} and some $\sim 10^{11}$ in the outer Oort cloud \citep[$a > 20~000~\rm{au}$, ][]{d775b1cc10ec4569af8397a8819ff9ba, 2013Icar..225...40B}. By scaling the population of km scale objects using a collisional cascade distribution which follows $N(d>D) \propto D^{-2.5}$ \citep{1969JGR....74.2531D}, we estimate the number of bodies with a diameter greater than 200 meters
\begin{equation}
\label{eqn: whole Oort}
N_{D>100\rm{m}} \sim 2.5 \times 10^{15}.
\end{equation}

The authors recognize that extrapolation along the same collisional cascade distribution ignores the variation of material strength with size \citep{2003Icar..164..334O}, but this order of magnitude estimate demonstrates the feasibility of unbound Oort clouds producing a significant population of interstellar objects.

If exo-Oort clouds resemble our own, and if some non-trivial fraction $f_o$ of every Oort cloud is lost to the ISM (we take a fiducial value of $f_o = 0.6$), then this mode of mass ejection alone could account for

\begin{equation}
\label{eqn: Oort loss}
n_{IS} \approx 0.10 (\frac{f_0}{0.6})~\rm{au}^{-3}.
\end{equation}

We find it inescapable that any Oort cloud like our own will be released at the end of a star's main-sequence lifetime. The number density of released objects approaches the one implied by the detection of \OBJECT. Thus, the progenitor of \OBJECT might be a white dwarf.

\section{Dryness}

\citet{engelhardt2017} calculated an upper limit on the interstellar number density of cometary bodies based on non-detections in Pan-STARRS, the Mt. Lemmon survey, and the Catalina Sky Survey. This limit is much lower what we have derived here because cometary bodies are much more visible at the same mass due to sublimation. ($2.4\times 10^{-4} \,\rm{au} ^{-3}$ at a confidence of 90\%). Comparing this upper limit on cometary bodies to our derived density of rocky bodies in \S\ref{sec:density}, the ratio of rocky to icy bodies in interstellar space must be in the realm of 

\begin{multline}
\label{eqn: dry}
f_{\rm{dry}} \gtrsim \frac{n_{rock}}{n_{rock}+n_{ice}} = \\ \frac{0.2\,\rm{au}^{-3}}{0.2+ 2.4\times 10^{-4}\,\rm{au}^{-3}} = 0.9988 .
\end{multline}

Long period comets demonstrate that the Oort cloud does have icy materials, however this does not imply the Oort cloud is made only of icy materials. Manx objects (tailless comets) point to the existence of inner solar system material that was scattered to the Oort cloud \citep{meech2016}. Without the typical cometary activity of long period comets, these Manx objects are difficult to detect, but not impossible. Pan-STARRS and other all-sky surveys report roughly a dozen per year \citep{2017EPSC...11..582H}. The ratio of rocky to icy material in the Oort cloud is thought to be $\sim 10^{-4} - 10^{-2}$ depending on the model \citep{2011Natur.475..206W, 2013ApJ...767...54I, 2015MNRAS.446.2059S, 1997ApJ...488L.133W, 2015Natur.524..322L}. Therefore Oort cloud objects cannot become interstellar objects directly without some surface drying mechanism.

The lack of a detected icy interstellar body implies that these unbound Oort cloud objects must have developed a crust thick enough to limit the sublimation of volatiles to match the ratio of rocky to icy bodies derived in equation \ref{eqn: dry}. \citet{2017arXiv171105687J} introduced the concept of cosmic ray exposure forming a crust on the surface of \OBJECT and subsequently expanded upon by \citet{Fitzsimmons} which included supernovae as a possible mechanism for surface volatile removal. Indeed, if supernovae or extended cosmnic ray exposure can form a crust on the surface of these liberated Oort cloud objects thick enough to prevent sublimation of volatiles at a few au from the Sun, this could be a direct source of dry, interstellar bodies.

Yet the issue with drying interstellar objects through these mechanisms is timing. Unless \OBJECT is much older than our solar system, it has experienced roughly the same amount of cosmic ray exposure as the active comets in our solar system. If cosmic rays removed all of the volatiles from \OBJECT, we would expect the same of these comets, yet this is observationally not the case. Instead, \OBJECT may have been within ``frying'' distance of a supernova. An object located $10^4$~au from a Type II supernova releasing $\sim 10^9 L_\odot$ over 10 days experiences the same amount of incident energy as 1~Myr of cosmic ray exposure. In clusters, a single supernova can cook exo-Oort clouds of multiple stars simultaneously, perhaps creating a reservoir of future dry, insterstellar objects. 

\section{Conclusion}
\label{sec:conclusion}

We have presented a thorough calculation of the number density of interstellar objects implied by the detection of \OBJECT by Pan-STARRS. Many aspects of \OBJECT are both intriguing and troubling, the derived number density of $0.2~\rm{au}^{-3}$ implies there are likely several of these objects in the inner Solar System at any given time. Comparing the cross section of Pan-STARRS to that of the Earth, we should expect one of these interstellar interlopers to hit the Earth every $\sim 30 \,\rm{Myr}$.

We have shown that the odds of detecting an object on an incoming orbit are significantly lower than detecting one from the opposite side of the Sun due to gravitational focusing effects. Thus, we should expect that the majority of interstellar objects will have passed perihelion by the time they are observed. Additionally, \OBJECT is roughly the size of most likely detection for Pan-STARRS. Our calculations make generous assumptions about Pan-STARRS' survey capability potentially resulting in an overestimated survey volume and an underestimated interstellar density.

The final issue regarding \OBJECT like objects is the mass ejection rates needed to litter the ISM with them. Dry, rocky bodies can be ejected during planet formation, but we have shown that the rate of mass ejection exceeds current predictions of inner solar system planetesimal disk mass, at least for Sun like stars. Unless the Sun has very few planetesimals compared to other stars, integrating across the range of stellar masses is unlikely to help.

Additionally, an alternative method of mass ejection, Oort cloud freeing, may be able to produce the requisite ISM number densities. However, this ejection source requires some crust-generation mechanism since observational limits on interstellar cometary objects are quite stringent. Objects like \OBJECT are 1~000 times more common than icy ones. If further study of the Oort cloud reveals a wealth of small rocky bodies this may not be an issue.

With only one detection any inferred rate carries intrinsic uncertainty. Perhaps the arrival of \OBJECT was a fluke, perhaps these interstellar objects do not have a homogeneous and isotropic distribution (``interstellar meteor showers''), or perhaps this is simply the first detection of many. Regardless, any hypothesis concerning the source of interstellar objects must satisfy Eq. \ref{eq:m_ej} and explain the dry surface of \OBJECT. The next detection of an interstellar interloper will shed more light on the nature of such objects in the ISM and their progenitors.

\section*{Acknowledgements}
We thank Karen Meech, Ben Shappee, Nader Haghighipour, and Jonathan Williams for multiple useful conversations.

\bibliographystyle{aasjournal.bst}
\bibliography{bibliography}

\end{document}